\documentclass{moriond}

\def\be{\begin{equation}}
\def\ee{\end{equation}}
\def\bea{\begin{eqnarray}}
\def\eea{\end{eqnarray}}

\newcommand{\Photo}{\includegraphics[height=35mm]{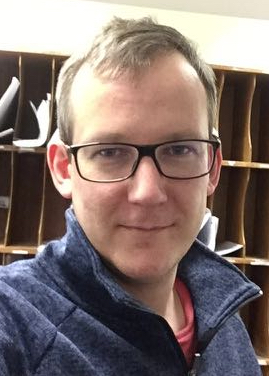}}

\begin{document}
\vspace*{2cm}
\title{CMB foregrounds - A brief overview}

\author{Clive Dickinson}

\address{Jodrell Bank Centre for Astrophysics, School of Physics \& Astronomy, Alan Turing Building, \\
The University of Manchester, Oxford Road, Manchester, M13 9PL, U.K.}

\maketitle
\abstracts{
CMB foregrounds consist of all radiation between the surface of last scattering and the detectors, which can interfere with the cosmological interpretation of CMB data. Fortunately, in temperature (intensity), even though the foregrounds are complex they can relatively easily be mitigated. However, in polarization, diffuse Galactic radiation (synchrotron and thermal dust) can be polarized at a level of $\ge 10\,\%$ making it more of a challenge. In particular, CMB B-modes, which are a smoking-gun signature of inflation, will be dominated by foregrounds. Component separation will therefore be critical for future CMB polarization missions, requiring many channels covering a wide range of frequencies, to ensure that foreground modelling errors are minimised.
}


\section{Introduction}

The Cosmic Microwave Background (CMB) radiation is the most distant source of electromagnetic radiation in our Universe, at a redshift $z \approx 1100$.\cite{Bennett2003} The CMB is therefore a backlight to all other sources of radiation between the surface of last scattering and the observer, which can contaminate the primordial signal - these sources of contamination are known as ``CMB foregrounds". To accurately measure the CMB and its anisotropies, foregrounds must be removed either by a component separation algorithm or marginalised over during power spectrum and/or cosmological parameter estimation.\cite{Planck2015_IX,Planck2015_X,Planck2015_XI} These methods rely on a prior knowledge of the foregrounds, such as intensity/polarization maps, statistical information on their spatial fluctuations, and most importantly, spectral information.\cite{Leach2008,Dunkley2009} Indeed, component separation has become a field in itself with applications in other areas of astronomy such as Epoch of Reionization and Intensity Mapping. The increasing difficulty of dealing with foregrounds, particularly in polarization, has been highlighted by the initial detection of large-scale B-modes from the BICEP2 collaboration, which has been shown to be largely due to thermal dust emission.\cite{BICEP2Planck}

In this article, I briefly review the sources of foreground contamination, their basic characteristics, and comment on the difficulties for future CMB missions, particularly for inflationary B-modes on large scales.

\setlength{\tabcolsep}{4pt}
\begin{table}[t]
\caption[]{Examples of the various types of CMB foregrounds. Some of these can be mitigated by careful design of the experiment, and in particular, by observing from a space satellite. Only a few astrophysical foregrounds are significantly polarized (``low" polarization typically means $<1\,\%$) - most notably, diffuse Galactic emission, which is polarized at $\ge 10\,\%$, and is dominant on large angular scales ($>1^{\circ}$).}
\label{tab:fg}
\begin{center}
\footnotesize
\begin{tabular}{ l | c |c | l }
\hline
Foreground 					&Polarization		&Angular scales    	&Mitigation technique  \\
\hline
Atmosphere  					&$\approx 0\,\%$     	&Large scales       	&Space/Balloon/high altitude sites \\
Ground          					&Varies          		&Large scales       	&Space/ground shield/low beam sidelobes \\
Radio Freq. Interference (RFI)  		&$0-100\,\%$  		&All     			&Space/remote sites/ground shield \\
Sun/Moon      					&Low              		&All                       	&Space/low beam sidelobes \\
Planets/solar system objects 		&Low   			&Small scales 		&Low frequencies / high resolution \\
Zodiacal light                                     	&Low			&Large scales		&Low frequencies    \\
Galactic synchrotron radiation          	&$\approx 10-40\,\%$ &Large scales		&Spectrum/high freqs.   \\
Galactic free-free radiation      		&Low                   	&Large scales		&Spectrum/H$\alpha$/recomb. lines \\
Galactic thermal dust radiation          	&$\approx 2-20\,\%$ &Large scales		&Spectrum/low freqs/starlight abs. \\
Galactic spinning dust radiation  	&Low			&Large scales		&Spectrum/FIR templates  \\
Galactic magnetic dust radiation	&$0-35\,\%$		&Large scales		&Spectrum \\
Galactic line emission (e.g. CO) 	&Low			&Large scales		&Narrow bandpasses  \\
Radio galaxies                                  	&Few \%                	&Small scales		&High frequencies/high resolution \\
Sub-mm/IR galaxies				&Low			&Small scales		&Low frequencies/high resolution  \\
Infrared Background (CIB)			&Low			&Small/interm. scales &Low frequencies/very high resolution \\
Secondary Anisotropies			&Low			&All				&Spectrum/spatial  \\				
\hline
\end{tabular}
\end{center}
\end{table}
\normalsize


\section{Review of foregrounds}

\begin{figure}
\begin{center}
\includegraphics[width=0.8\linewidth]{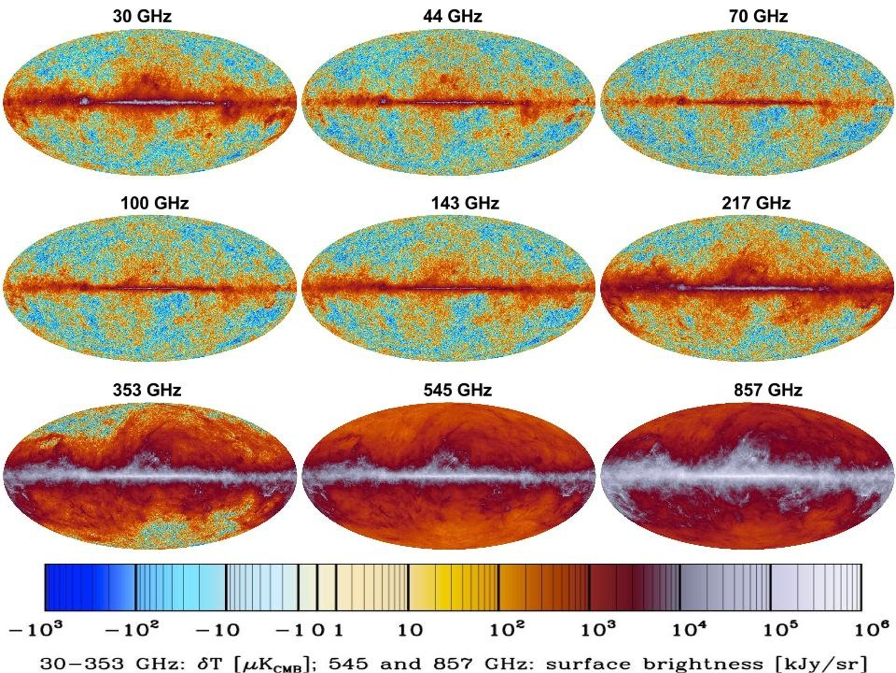}
\caption[]{{\it Planck} temperature maps, covering frequencies from 30\,GHz to 857\,GHz.\cite{Planck2015_I} The diffuse Galactic foregrounds are very visible across all the maps, but dominating at the lowest and highest frequencies. The foreground minimum is at $\approx 70$\,GHz; the exact value depends on angular scale and sky coverage.}
\label{fig:planckmaps}
\end{center}
\end{figure}

\begin{figure}
\begin{center}
\includegraphics[width=0.95\linewidth]{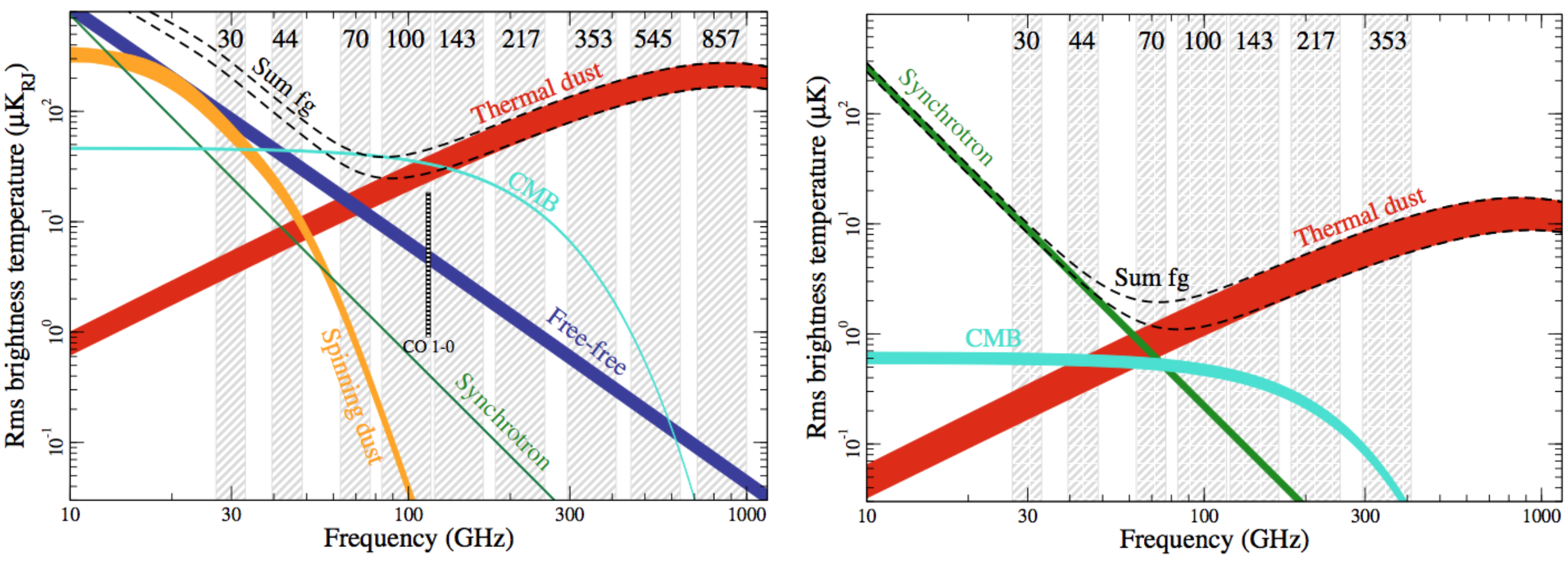}
\caption[]{Spectral characteristics of foregrounds and CMB in temperature ({\it left}) and polarization ({\it right}).\cite{Planck2015_X} These estimates represent the average fluctuations on $\approx 1^{\circ}$ scales over $\approx 80\,\%$ of the sky. The foregrounds in temperature appear to be more complex, but are comparable or below the CMB signal. In polarization, the number of significant components may be less, but their amplitudes are well above the typical level of CMB polarization anisotropies at all frequencies.}
\label{fig:spectrum}
\end{center}
\end{figure}

Table\,\ref{tab:fg} gives examples of common types of foregrounds, many of which can be mitigated by careful design of experiments. Fig.\,\ref{fig:planckmaps} shows the {\it Planck} temperature maps from 30\,GHz to 857\,GHz. The foregrounds are clearly visible, most notably from diffuse Galactic radiation, which dominate at the lowest and highest frequencies. Fig.\,\ref{fig:spectrum} shows the spectra of diffuse foregrounds in temperature and polarization, as derived from {\it Planck} data.\cite{Planck2015_X} There are at least 4 distinct foregrounds in temperature, each with a different spectral signature. In polarization, there are currently two known foregrounds, in the form of diffuse synchrotron and thermal dust emission.

In this review, we focus on diffuse Galactic radiation, which is the dominant foreground on large angular scales ($>1^{\circ}$) and in polarization.

\subsection{Synchrotron radiation}

Synchrotron radiation is emitted by relativistic cosmic ray (CR) electrons, which are accelerated by the Galactic magnetic field. In a regular magnetic field, the electrons spiral around the field lines, emitting radiation as they traverse a circular path. The emission depends on the number and energy spectrum of the CR electrons and the strength of the magnetic field. The spectrum of synchrotron radiation is directly dependent on the CR energy spectrum, which can vary across the sky, but is well-approximated by a power-law, at least over a range of frequencies. Most of the information on Galactic synchrotron radiation has come from low-frequency radio surveys. At frequencies above a few hundred MHz, the spectrum is optically thin and is steeply falling with frequency, with typical temperature spectral indices ($T \propto \nu^{\beta}$) of $\beta \approx -2.7$ at GHz frequencies, with variations $\Delta \beta \approx \pm 0.2$.\cite{Platania1998,Platania2003} At higher frequencies, the spectrum appears to steepen further, presumably due to radiative losses, which cause spectral ageing, with $\beta \approx -3.0$ at WMAP/{\it Planck} frequencies.\cite{Davies2006,Bennett2003,Planck2015_XI} However, multiple components along the line-of-sight can also flatten the spectrum.

\begin{figure}
\begin{center}
\includegraphics[width=0.9\linewidth]{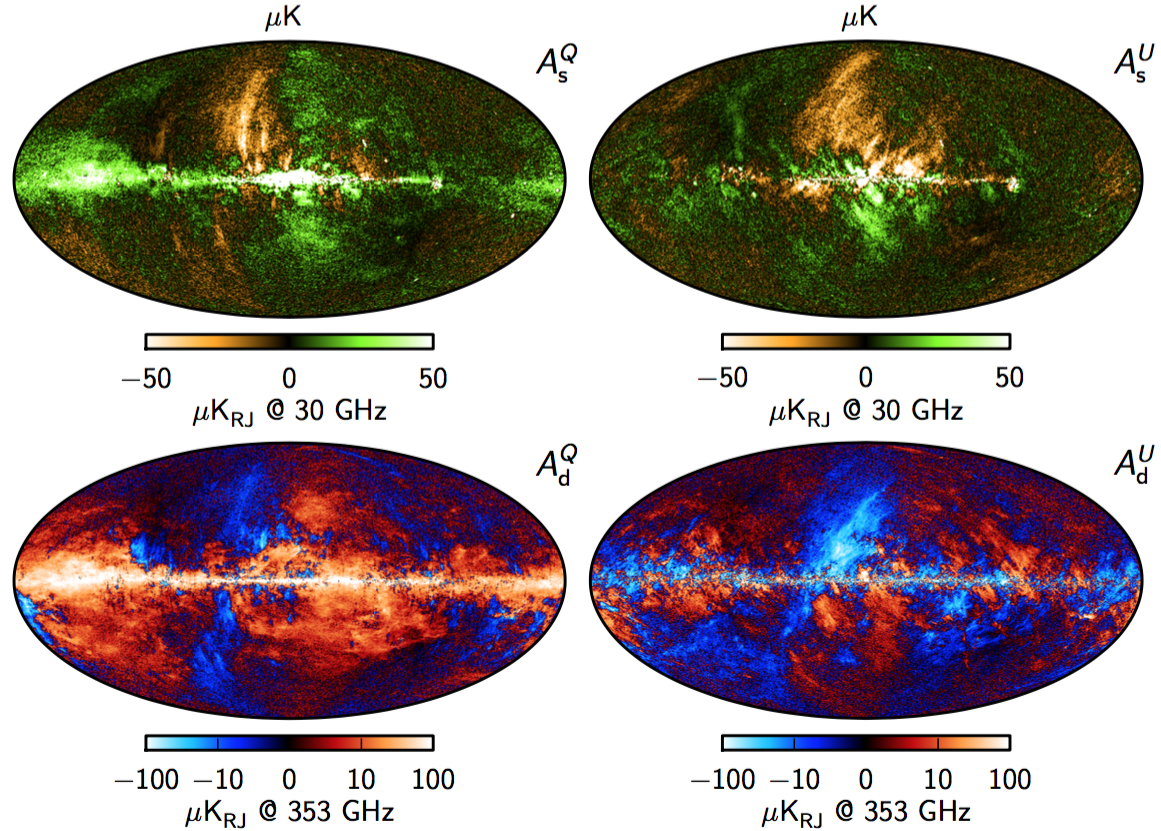}
\caption[]{Amplitude maps of polarization foregrounds in Stokes $Q$ ({\it left}) and $U$ ({\it right}) of synchrotron emission at 30\,GHz ({\it top}) and thermal dust emission at 353\,GHz ({\it bottom}). These maps are from the {\sc Commander} component separation analysis.\cite{Planck2015_X}}
\label{fig:fgpolmaps}
\end{center}
\end{figure}

The polarization of synchrotron radiation is less well known, since it can only be directly mapped at frequencies above a few GHz, due to the effect of Faraday Rotation, which is important at frequencies below a few GHz.\cite{Wolleben2006} The best maps to date are CMB data from WMAP and {\it Planck}, which are limited in S/N ratio and do not allow accurate spectral indices to be derived (except for relatively bright emission). Fig.\,\ref{fig:fgpolmaps} shows a model of the synchrotron polarization at 30\,GHz. Large-scale features are readily visible  and contribute significant polarization even at high Galactic latitudes. The polarization fraction of synchrotron radiation can, in principle, be as high as $75\,\%$, in a uniform and regular magnetic field. In practice, depolarization along the line-of-sight and non-regular fields reduces the polarized intensity. Even though we have maps of polarized synchrotron emission, the precise values of polarized fractions are difficult to quantify because the intensity of synchrotron emission is not well-known at frequencies above a few GHz due to the contributions of free-free and anomalous microwave emission. Nevertheless, at high latitudes, synchrotron emission is polarized at a level of $10-40\,\%$.\cite{Vidal2015,Planck2015_XXV}

\subsection{Free-free radiation}

Free-free radiation, or thermal bremsstrahlung, is emitted by free electrons interacting with ions, in ionised gas. As electrons are accelerated by ions (usually protons) they emit free-free radiation. Free-free radiation has a well-understood spectrum. At frequencies above a few GHz, where the emission is optically thin, the spectrum has a temperature spectral index of $\beta=-2.1$, with very little variation with electron temperature.\cite{Dickinson2003} At higher frequencies, approaching 100\,GHz, the spectrum steepens slightly to $\beta=-2.13$.\cite{Planck_int_XV} The relatively flat spectral index means that it could be the dominant foreground near the foreground minimum ($\nu \approx 70$\,GHz). The confusion with synchrotron, AME, and thermal dust, means it can be difficult to separate accurately. However, the close correlation with the optical H$\alpha$ line allows the amplitude to be derived independently.\cite{Dickinson2003} At high Galactic latitudes, free-free emission is relatively weak with $\Delta T \approx 10\,\mu$K at 30\,GHz, or $\Delta T \approx 1\,\mu$K at 100\,GHz.

Free-free is intrinsically unpolarized. Coulombic interactions are by their nature random in orientation, with no significant alignment with the magnetic field. Residual polarization can occur on sharp edges due to Thomson scattering. At high Galactic latitudes, the polarization is therefore expected to be very low ($\ll1\,\%$), with current measured upper limits at $<3\,\%$ \cite{Macellari2011} for diffuse emission and $\ll1\,\%$ for compact HII regions.  Free-free emission is probably not going to be a major foreground for CMB polarization studies.

\subsection{Thermal dust radiation}

Thermal dust radiation is blackbody emission modified by opacity effects, from interstellar dust grains with typical temperatures $T \approx 20$\,K. The observed spectrum is often modelled as a modified black-body, $T(\nu)=\tau \nu^{\beta_{d}} B(\nu,T_{d})$, where the $\nu^{\beta}$ term represents the change in emissivity of the grains with wavelength. $\beta_d$ is referred to as the emissivity index and is generally in the range $\beta=1-2$ although it can be steeper. {\it Planck} data, combined with IRAS/COBE data, have provided full-sky measurements of the thermal dust spectrum.\cite{Planck2013_XI} The thermal dust emission is reasonably well-modelled by a single modified-blackbody with mean temperature $T_d \approx 19$\,K and index $\beta_d \approx 1.6$.\cite{Planck_int_XLVIII} This value is somewhat flatter than had previously been predicted ($\beta\approx 1.7-2.0$), making it more difficult to observe the CMB at frequencies $\approx 100$\,GHz.

\begin{figure}
\begin{center}
\includegraphics[width=0.6\linewidth]{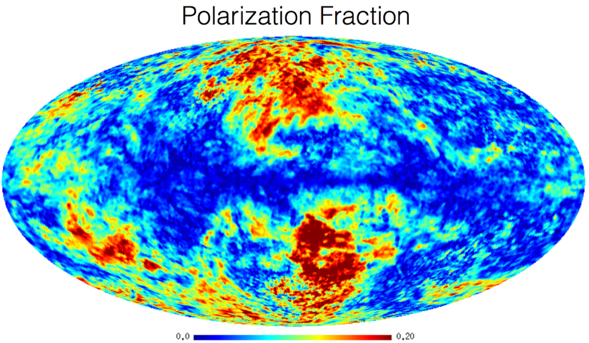}
\includegraphics[width=0.35\linewidth]{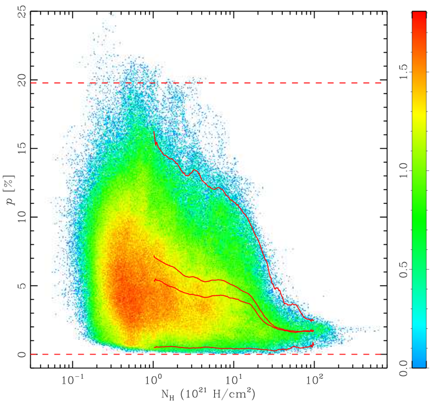}
\caption[]{{\it Left}: Full-sky map of thermal dust polarization fraction, as seen by {\it Planck} at 353\,GHz.\cite{Planck_int_XIX} {\it Right}: Polarization fraction of thermal dust emission, as seen by {\it Planck} at 353\,GHz, against an estimate of the total column density of the interstellar medium along the line-of-sight (which is essentially a scaled version of the intensity at 353\,GHz).\cite{Planck_int_XIX}}
\label{fig:thermaldustpol}
\end{center}
\end{figure}

Thermal dust emission can be significantly polarized. Elongated dust grains emit preferentially along their shortest axes while large dust grains, which are important at sub-mm wavelengths, can align efficiently by the Galactic magnetic field, causing a net polarization. The mean polarization fraction at high latitude is significantly higher than initial estimates and measurements from the Archeops experiment, which were at $\approx  5\,\%$. Recent {\it Planck} data have allowed the polarization fraction to be measured across the sky, and has found polarization fractions of up to $\approx 20\,\%$ with a mean value at high latitude of $\approx 10\,\%$.\cite{Planck_int_XIX,Planck2016_XXXVIII} Fig.\,\ref{fig:thermaldustpol} shows estimates of the thermal dust polarization fraction across the sky, and as a function of the total column density. The latter was estimated from the optical depth at 353\,GHz, which, to first order, is simply the brightness at 353\,GHz. It can be seen that a large fraction of the sky has polarization fractions $>5\,\%$. More worryingly for CMB studies is that the higher polarization fractions tend to occur in regions of low column density, which means low intensity. Higher density regions, including the Galactic plane, tend to have lower polarization fractions due to the effect of line-of-sight depolarization. The power spectrum of thermal dust anisotropies appears to follow a power-law with a slope $\alpha \approx -0.4$,\cite{Planck2016_XXX} leading to larger fluctuations on large angular scales.

\subsection{Spinning dust radiation}
\label{sec:spinningdust}

Spinning dust radiation is emitted by the smallest ($\sim 10^{-9}$\,m) interstellar dust grains and molecules, which can rotate at GHz frequencies. If they have an electric dipole moment, they emit by electric dipole radiation.\cite{Draine1998} The spectrum of spinning dust is invariably a highly peaked spectrum, which is dominated by the emission of the smallest grains (since they can spin the fastest due to lower moments-of-inertia). At least in a few Galactic clouds, observations have shown a good fit to a spinning dust spectrum with a typical peak frequency of $\approx 30$\,GHz. For the diffuse emission at high Galactic latitudes, the spectrum of spinning dust has not been measured due to the difficulty of component separation.

Theoretical work suggests that spinning dust is not highly polarized at frequencies above a few GHz.\cite{Lazarian2000,Hoang2016,Draine2016} Measurements of the level of polarization of spinning dust, largely consist of upper limits at the level of a few per cent.\cite{Dickinson2011,Rubino-Martin2012} This is one of the major arguments that AME is indeed due to spinning dust, and not magnetic dust (\S\,\ref{sec:magneticdust}). If all AME is due to spinning dust, then even a low ($\approx 1\,\%$) level of polarization could be problematic for future CMB data.\cite{Remazeilles2016} This is because of the very low target on $r$ and the unprecedented sensitivity of large detector arrays.

\subsection{Magnetic dust radiation}
\label{sec:magneticdust}

Magnetic dust (MD) radiation is emitted by interstellar dust grains and/or free-floating ferromagnetic material (e.g., Fe). Thermal fluctuations can change the net magnetization of a collect of grains by perturbing the alignment of spins in the magnetic domain, producing MD radiation.\cite{Draine1999} The spectrum of MD is similar to that of thermal dust emission, similar to a blackbody with temperature of order tens of K, depending on the type of magnetic materials and whether they are free-fliers or inclusions within dust grains. This is the main reason why MD emission is difficult to detect unambiguously. The most distinct feature(s) of MD radiation is that it can be highly polarized, and with a potentially distinct polarization signature as a function of frequency. A wide range of signatures can be produced depending on the precise model. The first works\,\cite{Draine1999} predicted, at least for a single-domain magnetic grains a $30-35\,\%$ polarization fraction at frequencies of $>200$\,GHz, falling to zero with lower frequency and then increasing again. More recent theoretical work\,\cite{Hoang2016} predicts a maximum polarization fraction of $10-15\,\%$ at high ($>200$\,GHz) and low ($<10$\,GHz) with a sharp transition somewhere in between.

MD radiation has not been definitively detected, although there have been a number of hints. The spectrum of the SMC has been shown to be readily fitted by including a plausible component of Fe particles, without which results in a poorer fit.\cite{Draine2012} {\it Planck} data have shown that the thermal dust spectrum appears to flatten at lower frequencies ($<350$\,GHz), which if not taken into account, results in a small excess at frequencies near 100\,GHz in intensity. Furthermore, the polarization fraction of the thermal dust emission correlated with thermal emission at 353\,GHz, appears to reduce slightly from 353\,GHz to 100\,GHz and below.\cite{Planck_int_XXII} These may be hints that a contribution of MD radiation is detectable in {\it Planck data}, although both of these can be explained by changes in dust grain properties, which affect the thermal dust emission. This complication of the spectrum near frequencies of 100\,GHz would clearly be problematic for CMB component separation in polarization.


\section{Measuring CMB B-modes}

A large fraction of the basic statistical information from the CMB has been extracted already. Although there are still several avenues to pursue (small-scales, SZ, large-scale anomalies, E-modes, lensing etc.), a major goal for the future is to constrain, and if possible detect, large-scale B-modes. The detection of primordial B-modes on scales $\ell < 100$ ($>2^{\circ}$) would be a smoking-gun signature of inflation. Furthermore, the amplitude of these B-modes, given by the tensor-to-scalar ratio $r$, determines the energy scale of inflation. The major obstacles, besides raw sensitivity, are the calibration requirements and component separation. Indeed, several CMB polarization satellite concepts have been put forward (B-pol, PRISM, CORE and CORE+ from ESA, EPIC and PIXIE from NASA, and LiteBIRD  from JAXA), which have the sensitivity to detect inflationary B-modes at the level of $r \approx 10^{-3}$ or better.\cite{Errard2016,Remazeilles2016} To achieve this our knowledge of the foregrounds and performance of component separation algorithms will need to be improved.

\begin{figure}
\begin{center}
\includegraphics[width=0.8\linewidth]{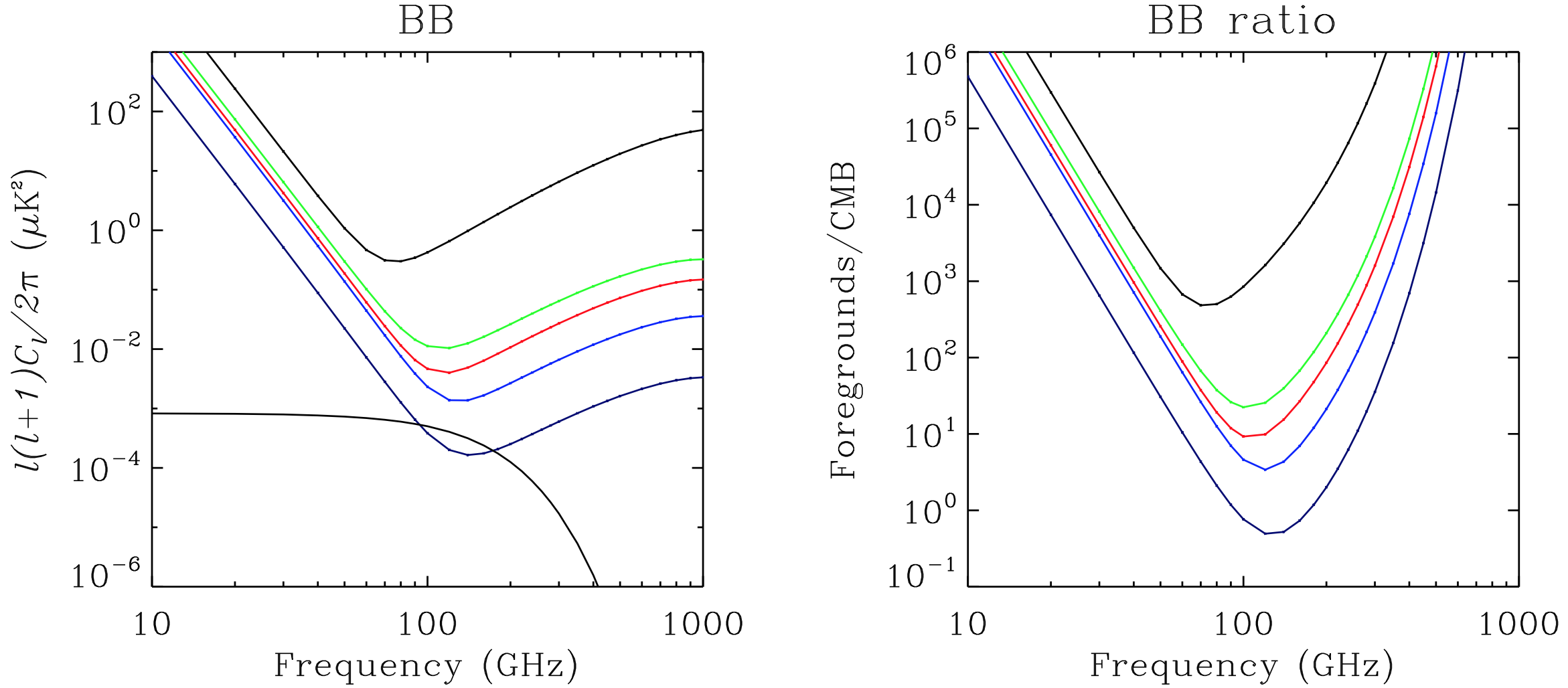}
\caption[]{Foregrounds B-mode power, as a function of frequency ({\it left}) and ratio of foregrounds-to-CMB B-mode power ({\it right}). The black curve represents the CMB B-mode power for $r=0.01$. The curves from top to bottom represent different sky areas (full-sky, 82\,\%, 50\,\%,23\,\%, and 5\,\%).\cite{Dunkley2009} The data were produced using the Planck Sky Model as of 2009. The low frequency side is accurate to within a factor of 2, primarily based on WMAP data. On the high frequency side, it assumes an average thermal dust polarization fraction of $\approx 5\,\%$ since it was made pre-{\it Planck}. New results from {\it Planck} show that the foreground power on the high frequency side is $\approx 2$ times larger in amplitude, or $\approx 4$ times in power.}
\label{fig:bmodesfg}
\end{center}
\end{figure}

Fig.\,\ref{fig:bmodesfg} depicts the level of diffuse foregrounds in B-mode polarization, for various sky areas, compared to the CMB, assuming $r=0.01$.\cite{Dunkley2009} It is clear that foregrounds dominate the CMB signal at all frequencies due to the fact that the foregrounds have a higher polarization fraction and that foregrounds emit similar amounts of E- and B-modes. It is important to note that these estimates were made pre-{\it Planck}, assuming a 5\,\% average polarization fraction for thermal dust. Therefore, in reality, the high frequency ($>100$\,GHz) foregrounds will have a factor of $\approx 2-4$ power. This clearly makes the situation worse, and is clearly born out by the BICEP2/Keck results. It also causes the foreground minimum to move to lower frequencies ($\approx 70\,$GHz).

The {\it Planck} data have revolutionised our knowledge of the microwave/sub-mm sky, particularly above 100\,GHz. However, the sensitivity of {\it Planck} polarization data, even at 353\,GHz (the highest frequency that contains polarization data, and therefore the highest signal-to-noise ratio), is still relatively low, particularly on scales $<1^{\circ}$. Future CMB missions will need to incorporate additional high frequency ($>300$\,GHz) channels as foreground monitors. This will not significantly affect the focal plane design since such detectors will take up a small amount of space because of their small size and only a small number of detectors would be required.

At low frequencies, CMB satellites will be limited by the focal plane area. Having additional low frequency channels uses up a lot of space (they will be physically large, scaling as $\lambda$ for a given angular resolution since they will likely be diffraction limited), which would reduce the overall sensitivity of the higher frequency ``CMB channels". Satellites are unlikely to include frequencies below about 30\,GHz. It will therefore be crucial to make high-sensitivity, high-fidelity ground-based surveys. Table\,\ref{tab:lowfreqsurveys} lists some of the main low-frequency polarization surveys underway. Foreground surveys should ideally be full-sky, have high signal-to-noise ratios per beam, and have minimal systematic errors.

We note that lensing of E-modes to B-modes by large-scale structure is also a foreground to the inflationary B-modes on the largest angular scales. Since lensing B-modes have the same frequency spectrum as the CMB, other methods are required, either based on spatial correlations or from direct subtraction using high-resolution measurements. 

We also note that the Cosmic Infrared Background (CIB) might also be a polarized foreground. The CIB consists of a large number of individual galaxies,\cite{Planck2011_XVIII} which individually are expected to have a low level ($<1\,\%$) of polarization, one might think that on average, should average to $\approx 0\,\%$ if they are orientated in random directions (as one would expect). This is true for the mean but since we are interested in the {\it fluctuations} it is the {\it variance} of the polarized intensity that matters, which does not cancel out. 

Finally, we remark on emission lines that can contaminate the detector bands and cause systematic errors in the data. This was a significant problem for {\it Planck} HFI, which was contaminated by the strongest sub-mm lines of CO (e.g., $J=1\rightarrow0$ at 115\,GHz). Bandpasses will need to be carefully chosen to avoid at least the very brightest emission lines. Furthermore, wide bandwidths require colour corrections to take into account the finite bandwidth, which causes the effective frequency to vary with the colour (spectrum) of the source. For ultra-sensitive experiments, this could be a limiting factor.

\begin{table}[t]
\caption[]{Low-frequency ground-based surveys relevant to large-scale CMB polarized foregrounds.}
\label{tab:lowfreqsurveys}
\begin{center}
\small
\begin{tabular}{ l | c |c | c | l }
\hline
Survey 				&Frequency		&Angular resolution    	&Sky   		&Status \\
\hline	
GEM					&0.4,1.4,2.3,5,10	&$\approx 0.5$ (10\,GHz) 	&Full-sky 		&On-going  \\
S-PASS				&2.3				&0.1					&Southern sky	&Obs complete \\
C-BASS				&5				&0.75				&Full-sky		&N complete; S started \\
QUIJOTE				&11,13,17,19		&$\approx 1$ 			&Northern sky	&On-going \\						
\hline
\end{tabular}
\end{center}
\end{table}
\normalsize

\begin{figure}
\begin{center}
\includegraphics[width=0.45\linewidth]{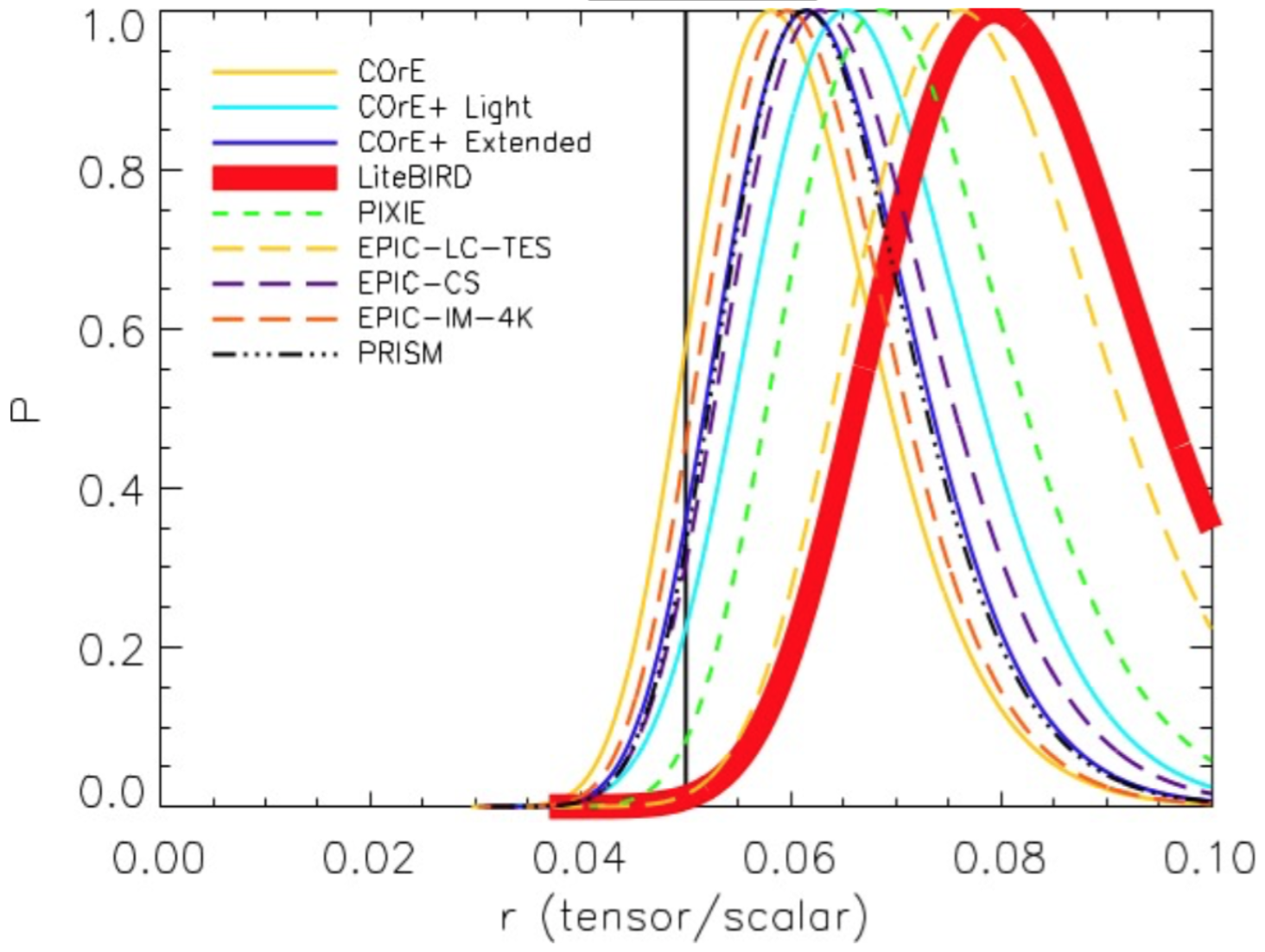}
\includegraphics[width=0.45\linewidth]{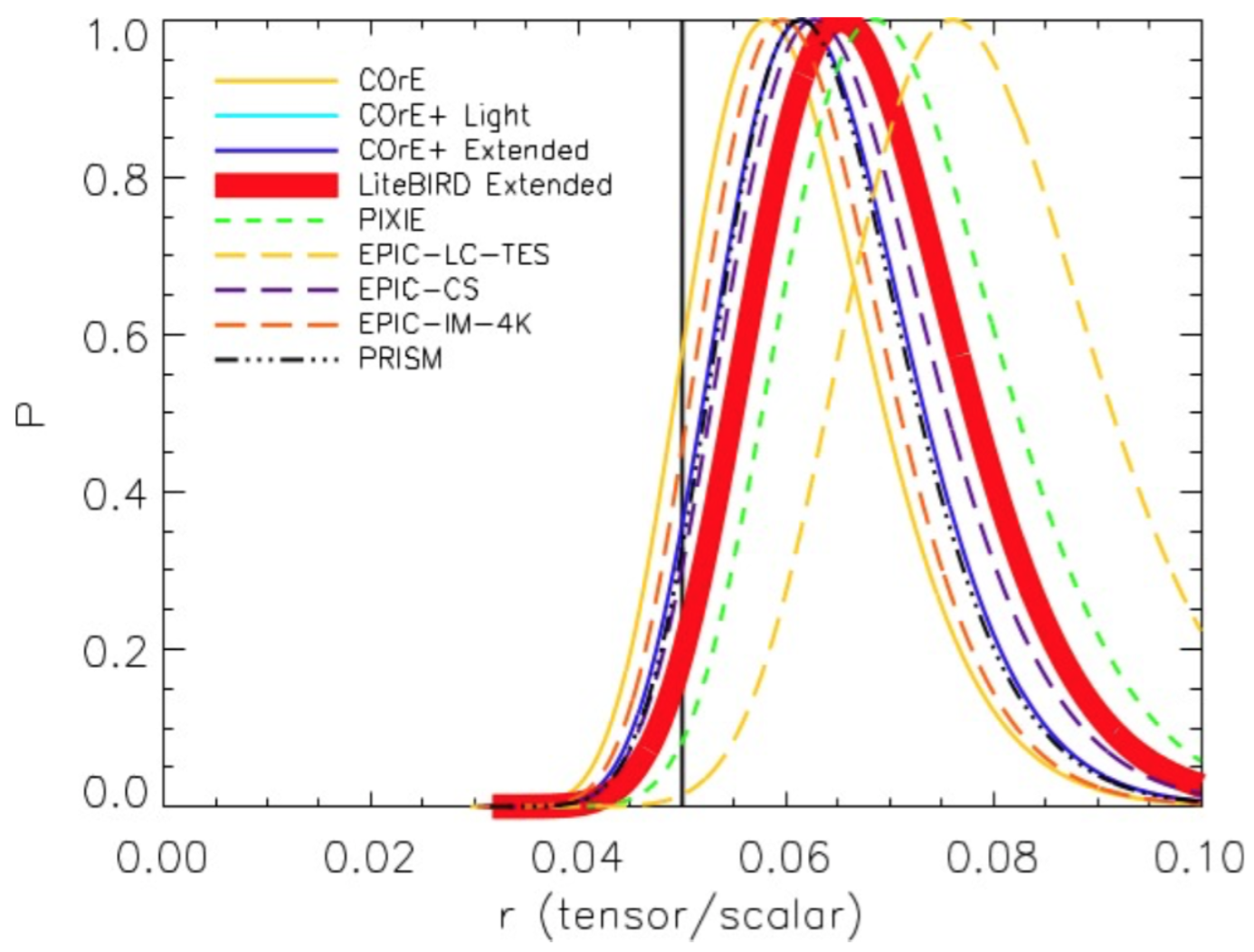}
\caption[]{Simulated posterior probability distributions for the tensor-to-scalar ratio, $r$, for a range of CMB polarization satellite mission concepts.\cite{Remazeilles2016} The input value is $r=0.05$. The foreground simulation includes synchrotron emission with positive curvature. The {\it left} panel shows that the results are biased high, due to the mis-modelling of the synchrotron foreground. The {\it right} panel shows how adding additional low frequency channels to the LiteBIRD experiment, the bias can be reduced.}
\label{fig:bpol}
\end{center}
\end{figure}

Fig.\,\ref{fig:bpol} shows the posterior probability distributions for $r$ for a range of CMB polarization satellite concepts. This simulation includes polarized thermal dust and synchrotron foregrounds, which were fitted for, using the {\sc Commander} code.\cite{Eriksen2008} The synchrotron spectrum includes positive curvature (flattening with increasing frequency), which causes a bias in the recovered estimates of $r$.  The LiteBIRD curve is most discrepant in the left-hand panel, due to its lack of frequency coverage. The right panel is the same but with additional frequency channels for the LiteBIRD configuration, which significantly reduces the bias. A bias in the recovered $r$-value can result either from incorrect foreground modelling or lack of frequencies to remove them.

\vspace{-1mm}
\section{Conclusions and outlook}

Foregrounds are a major issue for accurate studies of CMB anisotropies. Fortunately, in temperature, foregrounds appear not to be a major limitation. However, for detailed polarization studies, particularly large-scale B-modes, foregrounds are a more serious problem. It will therefore be vital for future missions to have as many frequency channels and as large a frequency range as possible. There are several proposed CMB polarization satellite mission concept, which typically have 6 or more frequency channels, up to frequencies of 350\,GHz and higher. However, at lower frequencies (below the minimum of foregrounds at $\approx 70$\,GHz) focal plane area limitations mean that satellite configurations can only afford one or two low frequency channels at most. It will therefore be critical to accurately measure the low frequency polarized foregrounds from ground-based telescopes operating at a few GHz up to $\approx 20$\,GHz. Component separation will likely be the limiting factor in achieving constraints on $r$ down to the $r=10^{-3}$ level and below. More frequency channels, covering a wide range of frequencies, will be critical in testing foreground mitigation.


\section*{Acknowledgments}
CD acknowledges support from an ERC Starting Consolidator Grant (no.~307209) and an STFC Consolidated Grant (ST/L000768/1). This review is dedicated to the late Prof.\,Rodney Deane Davies and Prof.\,Richard John Davis. They were both mentors and friends to me and many colleagues around the world.

\end{document}